\documentclass[preprint,10pt]{elsarticle}

\usepackage{amssymb}
\usepackage{amsmath}
\usepackage{booktabs}
\usepackage{microtype}
\usepackage{array}
\usepackage{graphicx}
\usepackage{placeins}
\usepackage{caption}
\usepackage{xcolor}
\usepackage{siunitx}
\usepackage{url}
\usepackage{orcidlink}
\hypersetup{hidelinks}

\journal{Preprint}

\newcommand{\mcc}{\mathrm{MCC}}
\newcommand{\hee}{H_{\mathrm{EE}}}

\begin{document}

\begin{frontmatter}

\title{Noise-Induced Predictability Redistribution Across Forecast Horizons of Extreme Events in Chaotic Dynamics}

\author[aff1]{Andrei Velichko\,\orcidlink{0000-0002-9341-1831}\corref{cor1}}
\ead{velichkogf@gmail.com}

\author[aff2]{Viet-Thanh Pham\,\orcidlink{0000-0001-5151-9812}}
\ead{phamvietthanh@iuh.edu.vn}

\cortext[cor1]{Corresponding author.}

\affiliation[aff1]{organization={Institute of Physics and Technology, Petrozavodsk State University},
            addressline={33 Lenin Ave.},
            city={Petrozavodsk},
            postcode={185910},
            country={Russia}}

\affiliation[aff2]{organization={Faculty of Electronics Technology, Industrial University of Ho Chi Minh City},
            city={Ho Chi Minh City},
            country={Vietnam}}

\begin{abstract}
Extreme events (EEs) in chaotic dynamics are rare broad excursions whose forecastability can be altered by dynamical noise. We investigate how noise changes EE occurrence and prediction skill across forecast horizons in a third-order autonomous chaotic flow. A single clean-data threshold is frozen for all realizations, broad events are defined by one maximum per excursion, and a future window $W=15$ is predicted from a 15-time-unit history using HistGradientBoosting with chronological data separation. As the forecast gap $G$ between the observed history and the future event window increases, the clean Matthews correlation coefficient (MCC) decreases from $0.641$ at $G=0$ to $0.165$ at $G=15$. Noise dependence is evaluated with ten paired realizations at eight amplitudes. The mean short-horizon score increases from $0.456$ in clean data to $0.546$ at $\sigma=0.007$; the paired gain is $0.0895$ (95\% CI $0.0494$--$0.1295$; Holm-adjusted $p=0.0234$). Noise strongly increases EE occurrence while event amplitude and width remain comparatively stable. Equalizing positive training counts across noise levels substantially attenuates the short-horizon gain, whereas strong noise reduces intermediate-horizon skill. We term this horizon-dependent, nonuniform change in forecast skill \emph{noise-induced predictability redistribution} (NIPR).
\end{abstract}


\end{frontmatter}

\section{Introduction}
\label{sec:intro}

Extreme events (EEs) are rare, large-amplitude departures from the typical behavior of nonlinear systems and arise in settings ranging from optical and neuronal dynamics to coupled networks, turbulence, climate-related models, and engineered oscillators. Their mechanisms include crises, intermittency, unstable invariant structures, multistability, and external perturbations, and the same observed extreme amplitude can be reached through dynamically different routes \cite{Chowdhury2022,Farazmand2019,Mishra2020,Cavalcante2013}. The combination of rarity, sensitivity to initial conditions, and nonlinear transient organization makes EEs simultaneously important and difficult to forecast. In this setting, predictability is not only a property of the classifier or regression model: it also depends on how an event is defined, which precursor information is observed, and how far ahead the forecast is required.

The literature on EE prediction has consequently evolved from generic warning indicators toward event-specific precursors and data-driven forecasting. Local-state and analogue ideas established that short chaotic histories can carry useful predictive information \cite{Farmer1987,Platzer2020}. Later studies formulated EE forecasting through physically motivated precursors, rare-event classification, reduced-order models, recurrent neural networks, and reservoir computing \cite{Bialonski2015,Blonigan2018,Guth2019,ChenMajda2020,Meiyazhagan2021,Mishra2022,Doan2021,Pammi2023,Ahmed2024,Wang2024,Martin2025}. Active-learning and neural-operator approaches have also been proposed for discovering and forecasting extreme responses \cite{Pickering2022}, while reservoir-computing studies have demonstrated that event-focused prediction can remain useful for several characteristic dynamical times in selected systems \cite{Doan2021,Pammi2023,Ahmed2024}. These results make clear that the relevant question is not merely whether a chaotic trajectory can be extrapolated, but whether the information contained in a finite past history remains discriminative for a specific future event.

This event-focused view naturally leads to the concept of a predictability horizon. Information-theoretic analysis shows that EE forecasting from observations has irreducible error arising from uncertainty in initial conditions, hidden variables, and model inadequacy \cite{Yuan2024}. In turbulent flows, events of similar amplitude can have markedly different predictability because they emerge from different large-scale patterns or causal routes \cite{VelaMartin2024JFM,VelaMartin2024PRF}. Very recent work further argues for a hierarchy of event-specific horizons and for instability-informed precursors that can extend practical warning times \cite{Yang2026,Katsidoniotaki2026,Consonni2026}. These developments motivate resolving forecast skill explicitly as a function of lead time rather than reporting a single aggregate accuracy.

Noise introduces a second layer of complexity. In conventional chaotic-state forecasting, observation noise usually decreases trajectory accuracy and shortens the usable prediction horizon \cite{Sangiorgio2021Sensitivity,Sangiorgio2021Forecasting,Vlachas2018}. Yet nonlinear systems also provide well-established examples in which finite noise is not purely detrimental. Coherence resonance shows that an intermediate noise level can maximize temporal organization in excitable dynamics \cite{Pikovsky1997}; stochastic forcing can improve regime predictability even while degrading trajectory predictability \cite{Kwasniok2014}; and an extended Lorenz--96-type system has exhibited stochastic-resonance-like forecasting improvement under external noise \cite{Revelli2010}. In machine-learning forecasting, noise added during training can act as a regularizer and, in suitable chaotic regimes, improve stability or predictive performance \cite{Zhai2023,Wikner2023,Vlachas2018}. These mechanisms are not interchangeable. In particular, noise injected into training data, observation noise, and dynamical noise acting directly on the evolution equations modify different parts of the forecasting problem. The present work concerns the last case.

Dynamical noise may also change the event-generation process itself. Noise-induced transitions can alter escape rates and residence times between metastable or coexisting dynamical states \cite{Forgoston2017}; additive perturbations modify recurrence and extreme-value statistics \cite{Faranda2012}; and in nonlinear oscillators and neuronal systems stochastic forcing can induce, suppress, or reorganize large-amplitude events \cite{Huang2022,Zhao2023,Hariharan2025,Boaretto2025,Sen2021}. Hence a change in forecast skill under noise need not imply that a fixed deterministic precursor has become intrinsically sharper. It may instead reflect altered event frequency, waiting-time statistics, or the number of positive precursor examples available to a data-driven classifier. This distinction is particularly important for rare-event learning, where changes in prevalence and positive-sample count can strongly affect training and operational metrics.

Taken together, these two bodies of literature suggest a distinction that has not been systematically resolved for broad EEs: noise can modify both the event process and the distribution of forecast skill over lead time. We use the term \emph{noise-induced predictability redistribution} (NIPR) for a nonuniform noise-induced change in EE forecast skill across forecast horizons. Operationally, relative to the paired clean baseline,
\begin{equation}
\Delta\mcc(G,\sigma)=\mcc(G,\sigma)-\mcc(G,0),
\label{eq:niprintro}
\end{equation}
and NIPR denotes a structured dependence of the sign or magnitude of $\Delta\mcc$ on both $G$ and $\sigma$, rather than a uniform vertical shift of predictive performance. Here, MCC denotes the Matthews correlation coefficient, a robust metric for binary classification that remains informative under class imbalance because it accounts jointly for true and false positives and negatives; the evaluation protocol is detailed in the Methods section. The term NIPR is deliberately descriptive: it does not presume classical stochastic resonance, nor does it imply that noise universally increases intrinsic predictability. Instead, it asks whether dynamical noise reallocates useful predictive information across lead times as the statistics of the extreme-event process change. In this work, we introduce the term NIPR for this horizon-resolved behavior. To the best of our knowledge, it has not previously been formulated explicitly as a mechanism describing how dynamical noise redistributes extreme-event predictability across forecast horizons.

The third-order autonomous chaotic flow considered here was recently studied numerically and experimentally as an EE-generating system \cite{Durairaj2026}. Its observable $y(t)$ contains broad extreme excursions with faster local oscillations embedded inside them. This structure exposes a methodological ambiguity that is often hidden by a simple threshold-crossing definition: treating every micropeak or every short threshold-occupancy interval as an independent target can alter event duration, prevalence, and the inferred forecast horizon. We therefore first define one physical broad event, then ask whether its broad maximum will occur within a future window. The same event definition and the same threshold are retained under noise.

The main contributions of this work are as follows. First, we formulate a single-threshold broad-event representation in which one coordinate is assigned to each large excursion and its half-height width is measured independently of the threshold. Second, we determine the clean-system history scale and finite-window predictability horizon with strict chronological train/validation/test separation. Third, after an exploratory dense noise scan, we perform a paired ensemble analysis using ten independent realizations. Within each replicate, every noise amplitude is compared with the corresponding clean trajectory generated from the same small initial-condition perturbation, while the same Gaussian-noise sequence is rescaled across the nonzero noise amplitudes. Fourth, because increasing dynamical noise also increases the number of EEs and therefore the number of positive training examples, we introduce a positive-training-count-matched control. For each paired realization and forecast gap, all noise levels are trained using the same number of positive examples, while validation and test data retain their natural event prevalence. This control assesses how much of the apparent noise-assisted prediction gain can be attributed simply to the larger number of EE examples available for training. Fifth, we identify and characterize the NIPR effect through the full $\Delta\mcc(G,\sigma)$ map: intermediate dynamical noise concentrates predictive advantage at short lead times, whereas strong noise removes or reverses that advantage farther ahead. This final step links the predictive response to the simultaneous noise-induced reorganization of EE occurrence statistics and therefore distinguishes NIPR from a generic claim of noise-enhanced forecasting.

The remainder of the paper is organized as follows. Section~\ref{sec:methods} defines the chaotic system, frozen EE threshold, broad-event geometry, finite-window prediction target, chronological validation protocol, paired ensemble, count-matched control, and operational NIPR measure. Section~\ref{sec:results} presents the event statistics, clean predictability horizon, and the NIPR response across noise amplitudes and forecast gaps. Section~\ref{sec:discussion} interprets NIPR in relation to previous work on EE predictability, noise-assisted forecasting, and noise-induced changes in event occurrence, and discusses the limits of the present claim. Section~\ref{sec:conclusions} summarizes the main conclusions.

\section{Methods}
\label{sec:methods}

\subsection{Chaotic system, integration, and stochastic forcing}

We consider the third-order autonomous flow \cite{LiSprott2016,Durairaj2026}
\begin{align}
\dot{x} &= z(a-y), \\
\dot{y} &= z^2-by, \\
\dot{z} &= x-cz,
\label{eq:system}
\end{align}
with $a=6$, $b=0.072$, and $c=0.29$. The reference initial condition is $(x_0,y_0,z_0)=(0.1,1,1)$. The equations are integrated in double precision with a fixed-step fourth-order Runge--Kutta (RK4) method using $\Delta t=10^{-3}$. Each trajectory has duration $T=50{,}000$, and the first $100$ time units are discarded.

Dynamical noise is introduced only in the $y$ coordinate after each deterministic RK4 step,
\begin{equation}
y_{n+1}=y_{n+1}^{\rm RK4}+\sigma\sqrt{\Delta t}\,\xi_n,
\qquad \xi_n\sim\mathcal{N}(0,1).
\label{eq:noise}
\end{equation}
The exploratory dense scan uses
\begin{equation}
\begin{aligned}
\sigma\in\{&0,10^{-4},3\!\times\!10^{-4},5\!\times\!10^{-4},7\!\times\!10^{-4},10^{-3},
1.3\!\times\!10^{-3},1.6\!\times\!10^{-3},\\
&2\!\times\!10^{-3},2.5\!\times\!10^{-3},3\!\times\!10^{-3},4\!\times\!10^{-3},
5\!\times\!10^{-3},7\!\times\!10^{-3},10^{-2},3\!\times\!10^{-2}\}.
\end{aligned}
\label{eq:sigmagrid}
\end{equation}
The final paired ensemble uses the representative subset defined in Sec.~\ref{sec:ensemble}.

\subsection{Single working extreme-event threshold}

The observable used for EE analysis is $y(t)$, consistent with the reference study \cite{Durairaj2026}. For threshold estimation only, every tenth post-transient sample is used, giving an analysis step $\Delta t_a=0.01$. A Savitzky--Golay filter with window length $21$ analysis samples (0.21 time units) and polynomial order $3$ is applied before detecting local maxima with a minimum separation of $0.5$ time units.

Let $y_{\rm peak}$ denote this clean-system local-maximum population. We define one working threshold
\begin{equation}
\hee=\langle y_{\rm peak}\rangle+3\,\sigma_{y_{\rm peak}}.
\label{eq:threshold}
\end{equation}
For the reference clean trajectory, $10{,}874$ local maxima give
\begin{equation}
\langle y_{\rm peak}\rangle=7.779442,\qquad
\sigma_{y_{\rm peak}}=1.757544,
\end{equation}
and therefore
\begin{equation}
\hee=13.052073.
\label{eq:thresholdvalue}
\end{equation}
This numerical value is computed once and then frozen. It is not recalculated when noise or the initial condition changes the event rate.

\subsection{Broad-peak extraction and event geometry}
\label{sec:broadpeaks}

The large excursions of $y(t)$ are broad structures containing faster oscillatory micropeaks. To assign one event coordinate to one broad excursion, the $\Delta t_a=0.01$ signal is smoothed with a Gaussian kernel of temporal standard deviation $1.0$. Broad maxima are detected with a minimum separation of $10$ time units and a minimum prominence of $0.5$. For each smoothed broad maximum, the exact maximum of the original unsmoothed $\Delta t=10^{-3}$ signal is recovered inside a $\pm8$-time-unit neighborhood. The recovered time and amplitude are denoted by $t_p$ and $A_p$.

Peak width is defined independently of $\hee$. Let $\bar y_\sigma$ denote the complete post-transient mean of one realization and $y_{p,\mathrm{smooth}}$ the height of the associated smoothed broad maximum. The half-height level is
\begin{equation}
y_{1/2}=\bar y_\sigma+
\frac{y_{p,\mathrm{smooth}}-\bar y_\sigma}{2}.
\label{eq:halfheight}
\end{equation}
If $t_L$ and $t_R$ are the left and right crossings of the smoothed curve with this level,
\begin{equation}
W_{1/2}=t_R-t_L.
\label{eq:width}
\end{equation}
A broad peak is included in the EE catalog when
\begin{equation}
A_p\ge \hee.
\label{eq:eventcriterion}
\end{equation}
Thus, the threshold selects events but does not determine their width.

\begin{figure}[!htbp]
\centering
\includegraphics[width=0.96\linewidth]{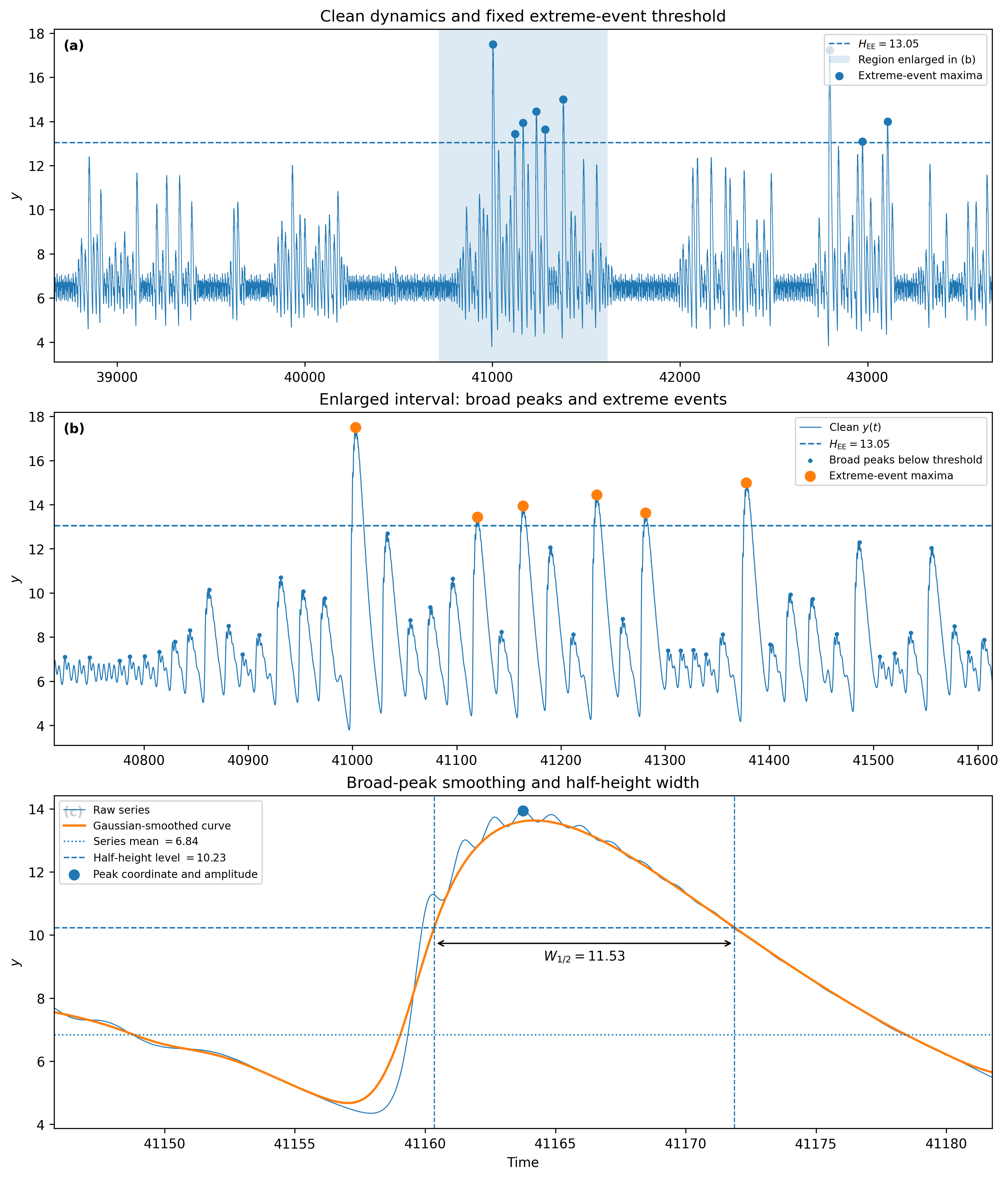}
\caption{Single-threshold broad-event definition. (a) Reference clean dynamics with the fixed threshold $\hee=13.0521$ and the interval enlarged in panel (b). (b) Broad maxima below the threshold and EE maxima above the same fixed threshold. (c) Raw and Gaussian-smoothed signals around one representative broad EE. The half-height level is measured relative to the mean of the complete realization, so $W_{1/2}$ is independent of the EE threshold.}
\label{fig:eventdefinition}
\end{figure}

\FloatBarrier
\subsection{Reference dense noise scan}

The same frozen $\hee$ and broad-peak extraction parameters are first applied to the $16$-level noise grid in Eq.~\eqref{eq:sigmagrid}. For each realization we record the number of selected EEs, event rate per $1000$ time units, peak-amplitude distribution, $W_{1/2}$ distribution, and intervals between consecutive EE maxima. This dense scan is used to identify the overall noise-response structure and the representative $\sigma$ range for the final ensemble; its single-realization statistics are not used as uncertainty estimates.

\begin{table}[!htbp]
\centering
\caption{Selected event statistics from the reference dense noise scan with the same frozen threshold $\hee=13.052073$.}
\label{tab:eventstats}
\small
\begin{tabular}{ccccc}
\toprule
$\sigma$ & $N_{\rm EE}$ & rate/1000 & median $A_p$ & median $W_{1/2}$ \\
\midrule
$0$ & 45 & 0.902 & 13.993 & 11.553 \\
$1.3\times10^{-3}$ & 62 & 1.242 & 14.361 & 11.733 \\
$3\times10^{-3}$ & 74 & 1.483 & 14.438 & 11.669 \\
$10^{-2}$ & 82 & 1.643 & 14.459 & 11.555 \\
$3\times10^{-2}$ & 101 & 2.024 & 14.168 & 11.299 \\
\bottomrule
\end{tabular}
\end{table}

\subsection{Broad-event finite-window forecasting}
\label{sec:predictionmethod}

For anchor time $t$, the forecast gap $G$ is the lead time between the end of the observed history and the beginning of the future event window. The binary target is defined directly from the broad EE peak catalog,
\begin{equation}
Y(t;G)=
\begin{cases}
1, & \exists\,t_p\in[t+G,t+G+W),\\
0, & \text{otherwise}.
\end{cases}
\label{eq:broadtarget}
\end{equation}
The future-window width is fixed at
\begin{equation}
W=15,
\label{eq:forecastwindow}
\end{equation}
which is slightly larger than the reference clean median $W_{1/2}=11.553$ and represents the physical broad-event scale. Threshold crossings within one excursion are not separate positive targets.

Prediction anchors are separated by one time unit. For clean history selection,
\begin{equation}
T_{\rm hist}\in\{15,30,45,60,75,90\}.
\label{eq:histories}
\end{equation}
The classifier input is represented by a fixed vector of $60$ feature positions. For the minimum history $T_{\rm hist}=15$, the first $30$ features are uniformly spaced samples from the most recent interval $[t-15,t]$, while the remaining $30$ positions are filled with zeros because no older history is available. For every longer candidate history, the first $30$ features are kept exactly the same, and the second block is replaced by $30$ uniformly spaced samples from the additional older interval $[t-T_{\rm hist},t-15)$. Thus increasing $T_{\rm hist}$ adds older information without changing the representation of the immediate $15$-time-unit precursor. The zero-filled block for $T_{\rm hist}=15$ is only a technical placeholder used to keep the input dimension fixed at $60$.

\subsection{Classifier, chronological validation, and clean history selection}

A \texttt{HistGradientBoostingClassifier} is used with learning rate $0.06$, $180$ boosting iterations, at most $15$ leaf nodes, minimum leaf size $15$, and $L_2$ regularization $0.2$. Positive anchors are rare, so the training set receives inverse-frequency balanced sample weights calculated separately for each forecast gap. No synthetic oversampling is used.

Each post-transient trajectory is split chronologically into $50\%$ training, $25\%$ validation, and $25\%$ test regions. History and future buffers prevent information or target windows from crossing split boundaries. A common future buffer equal to $\max(G)+W=45$ is used for all gaps. For each $(T_{\rm hist},G)$ in the clean selection stage, the classifier is fitted on training data, its probability threshold is chosen on validation data by maximizing MCC, and that threshold is applied unchanged to the test region. Candidate histories are compared by mean validation MCC over $G=0$--$15$; if two histories differ by less than $0.005$, the shorter history is preferred. The untouched test set is not used for history or probability-threshold selection.

The primary metric is the Matthews correlation coefficient (MCC), which uses all four entries of the confusion matrix and is therefore well suited to the class-imbalanced binary prediction problem posed by rare EEs. Average precision (AP), AP divided by event prevalence, ROC AUC, precision, recall, false-positive rate, and confusion counts are retained as secondary diagnostics. After selecting $T_{\rm hist}$, the forecast-gap grid is
\begin{equation}
G=0,1,2,\ldots,30.
\label{eq:ggrid}
\end{equation}

\subsection{Paired multi-realization noise validation and count-matched control}
\label{sec:ensemble}

The final noise analysis consists of ten independent paired replicates evaluated at
\begin{equation}
\sigma\in\{0,5\times10^{-4},2\times10^{-3},3\times10^{-3},
5\times10^{-3},7\times10^{-3},10^{-2},3\times10^{-2}\}.
\label{eq:ensemblesigma}
\end{equation}
The reduced grid in Eq.~\eqref{eq:ensemblesigma} is a representative subset of the broader exploratory grid in Eq.~\eqref{eq:sigmagrid}. The dense scan was used to resolve the overall dependence of prediction skill on noise amplitude and to identify the weak-, intermediate-, and strong-noise regimes of interest. The subset in Eq.~\eqref{eq:ensemblesigma} was then used for the computationally more expensive paired-ensemble validation, with ten realizations generated at each selected noise level. Thus, Eq.~\eqref{eq:sigmagrid} serves for exploratory localization of the noise response, whereas Eq.~\eqref{eq:ensemblesigma} serves for statistical validation of its reproducibility. The selected values retain the clean reference, a weak-noise case, several intermediate amplitudes spanning the positive NIPR region identified in the dense scan, and a strong-noise case used to test the loss of longer-horizon predictability. The clean-selected $T_{\rm hist}$, $W$, feature representation, model hyperparameters, chronological split, and validation-threshold procedure are frozen across this ensemble.

For replicate $r$, only the initial $x$ coordinate is perturbed,
\begin{equation}
x_0^{(r)}=0.1+\delta_x^{(r)},
\end{equation}
where $\delta_x^{(r)}$ takes ten symmetric values from $-4.5\times10^{-8}$ to $+4.5\times10^{-8}$ in steps of $10^{-8}$; $y_0=1$ and $z_0=1$ remain fixed. The same perturbed initial condition is used for all $\sigma$ within that replicate. A single Gaussian-noise seed is also assigned to each replicate and reset for every nonzero $\sigma$, so the underlying sequence $\xi_n^{(r)}$ is the same and only its amplitude $\sigma$ changes. The $\sigma=0$ trajectory contains no stochastic term but shares the same perturbed initial condition. This paired design reduces between-comparison variability while allowing independent replicate-to-replicate divergence.

The ensemble is generated in compact mode. The full raw $y(t)$ signal is kept only transiently in memory, broad events are extracted with the fixed Program-02 settings, and only the $\Delta t_a=0.01$ series is retained long enough for prediction. No ensemble raw trajectories are required for the final statistics.

For each replicate and $\sigma$, we summarize the horizon-dependent test skill as
\begin{align}
S_{\rm near}(\sigma) &= \frac{1}{6}\sum_{G=0}^{5}\mcc(G,\sigma), \label{eq:snear}\\
S_{\rm mid}(\sigma) &= \frac{1}{11}\sum_{G=5}^{15}\mcc(G,\sigma), \label{eq:smid}\\
S_{\rm full}(\sigma) &= \frac{1}{16}\sum_{G=0}^{15}\mcc(G,\sigma). \label{eq:sfull}
\end{align}

For each paired replicate $r$, we additionally compute the horizon-resolved change
\begin{equation}
\Delta\mcc_r(G,\sigma)=\mcc_r(G,\sigma)-\mcc_r(G,0),
\label{eq:niprmap}
\end{equation}
and average it across replicates to obtain the NIPR map $\langle\Delta\mcc(G,\sigma)\rangle_r$. In this work, NIPR is used as an operational description of a nonuniform response across $G$: a positive region denotes noise-assisted forecast skill relative to paired clean dynamics, a negative region denotes noise-induced degradation, and a horizon-dependent transition between these regions constitutes the redistribution signature. No separate resonance mechanism is assumed by this definition.

Because dynamical noise substantially increases EE occurrence, noisy trajectories generally contain more positive prediction anchors than the clean trajectory. An apparent improvement in forecast skill could therefore arise partly because the classifier is trained on a larger number of positive examples, rather than because the precursor dynamics has become intrinsically more informative. To assess this contribution, we additionally perform a positive-training-count-matched control. For each replicate $r$ and gap $G$, the common positive-training count is
\begin{equation}
N_{\rm match}(r,G)=
\min_{\sigma}N_{+}^{\rm train}(r,\sigma,G).
\label{eq:matchbudget}
\end{equation}
At every $\sigma$, positive training anchors are deterministically subsampled to $N_{\rm match}$ while all negative training anchors are retained. Importantly, only the training set is modified: validation and test sets remain completely unchanged and preserve the natural EE prevalence generated at each noise level. Thus, the matched analysis equalizes the amount of positive information available to the classifier during training without artificially modifying the prediction task at evaluation time.

Across the ten replicates, means, standard deviations, standard errors, medians, interquartile ranges, and two-sided $95\%$ Student-$t$ confidence intervals are computed. Noisy conditions are compared with the paired clean replicate using the Wilcoxon signed-rank test. Holm correction is applied across the seven nonzero $\sigma$ values separately for each summary score, and paired effect size $d_z$ is reported. Statistical significance is assessed using the Holm-adjusted $p<0.05$ criterion.

\section{Results}
\label{sec:results}

\subsection{The single threshold isolates broad extreme excursions}

The reference clean analysis yields $\hee=13.052073$ from $10{,}874$ local maxima. With this threshold and the broad-peak detector, the reference clean trajectory contains $45$ EEs over $49{,}900$ post-transient time units, corresponding to $0.902$ events per $1000$ time units. The clean median peak amplitude is $13.993$, and the median half-height width is $W_{1/2}=11.553$.

Figure~\ref{fig:eventdefinition} shows that the selected event is a broad structure extending over roughly a dozen time units rather than a sub-unit threshold-occupancy episode. The broad-peak coordinate $t_p$ therefore provides the physically relevant event marker for the prediction target.

\subsection{The dense reference scan shows stronger changes in occurrence than morphology}

\begin{figure}[!htbp]
\centering
\includegraphics[width=0.94\linewidth]{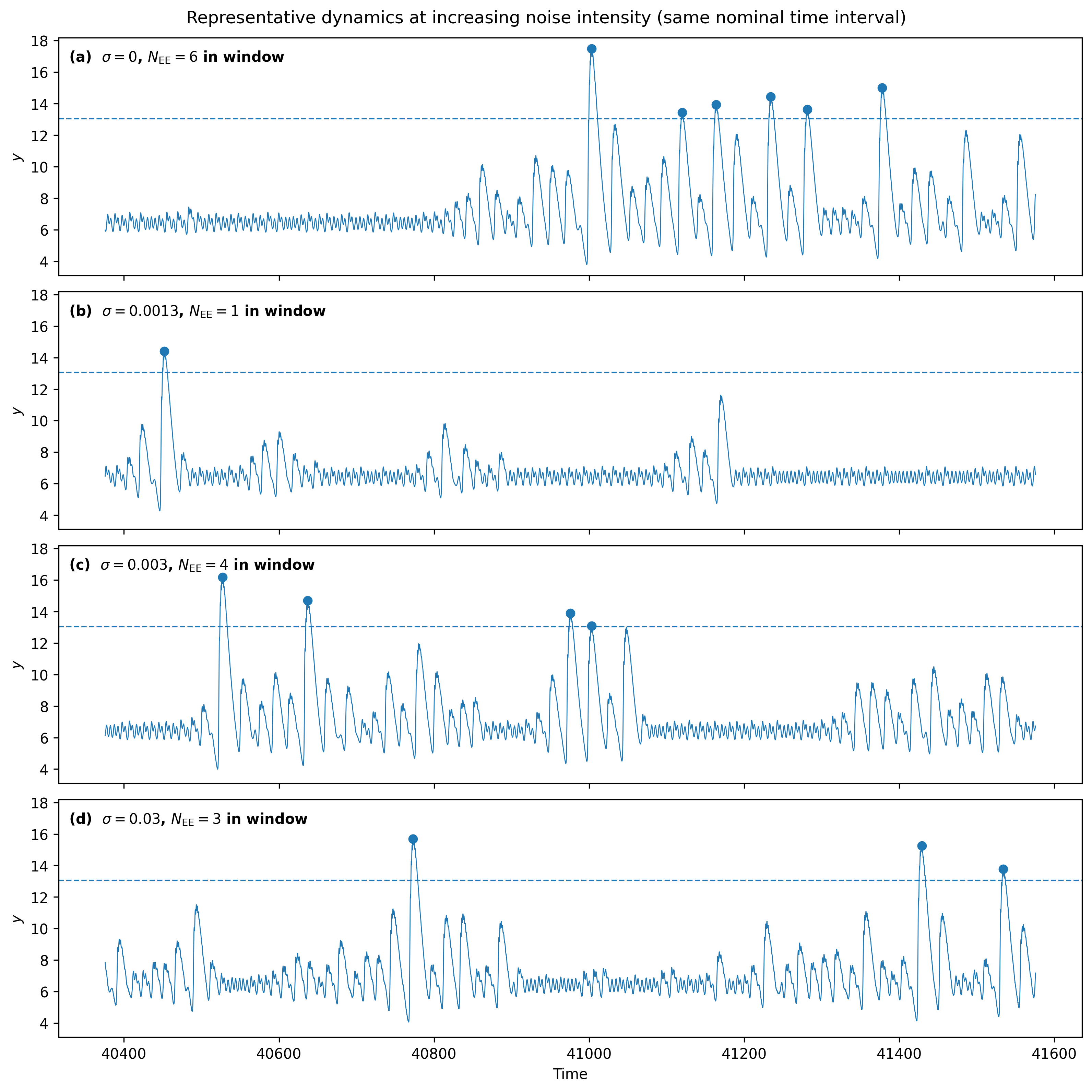}
\caption{Representative $y(t)$ segments for four noise intensities, plotted over the same nominal time interval with the same frozen $\hee$. Filled markers indicate selected broad EE maxima. The dynamical noise is small relative to the deterministic excursion scale and therefore does not produce obvious visual roughening; its main effect is revealed by long-time EE occurrence statistics.}
\label{fig:noisetraces}
\end{figure}

Figure~\ref{fig:noisetraces} shows representative clean and noisy trajectory segments using the same frozen $\hee$. Even at the largest noise amplitude considered here, the stochastic forcing is small relative to the deterministic excursion scale, so the noisy signal does not show visually obvious high-frequency roughening. This is expected because the perturbation acts dynamically at every integration step and changes the subsequent trajectory rather than being added afterward as observation noise. The broad excursion morphology therefore remains visually similar across the displayed cases; the principal effect of noise emerges statistically over long trajectories through changes in EE occurrence rate, event timing, and inter-event intervals.

\begin{figure}[!htbp]
\centering
\includegraphics[width=0.84\linewidth]{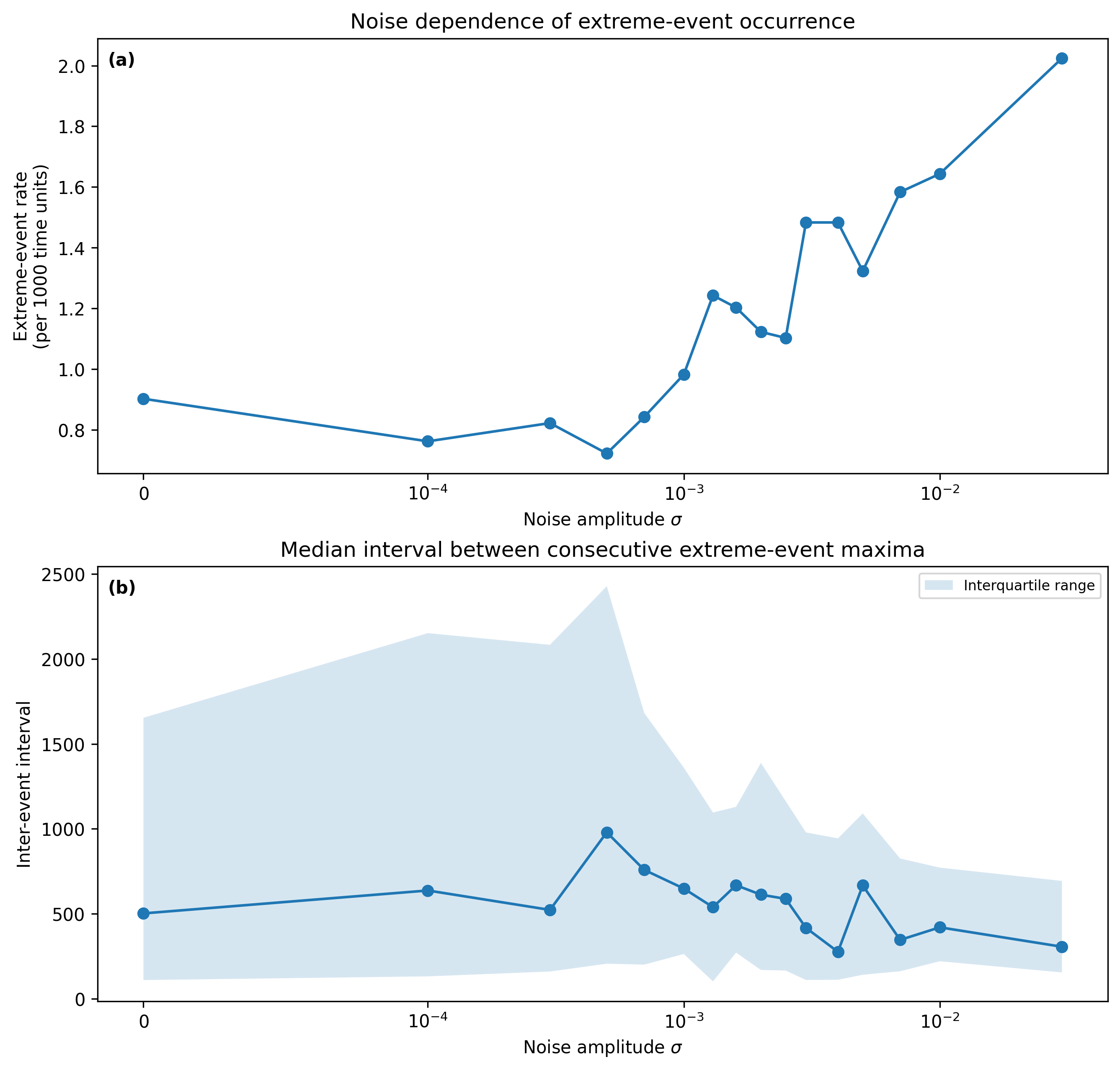}
\caption{Reference dense-scan EE occurrence statistics. (a) Event rate per $1000$ time units. (b) Median interval between consecutive EE maxima; shading shows the interquartile range.}
\label{fig:occurrence}
\end{figure}

\begin{figure}[!htbp]
\centering
\includegraphics[width=0.84\linewidth]{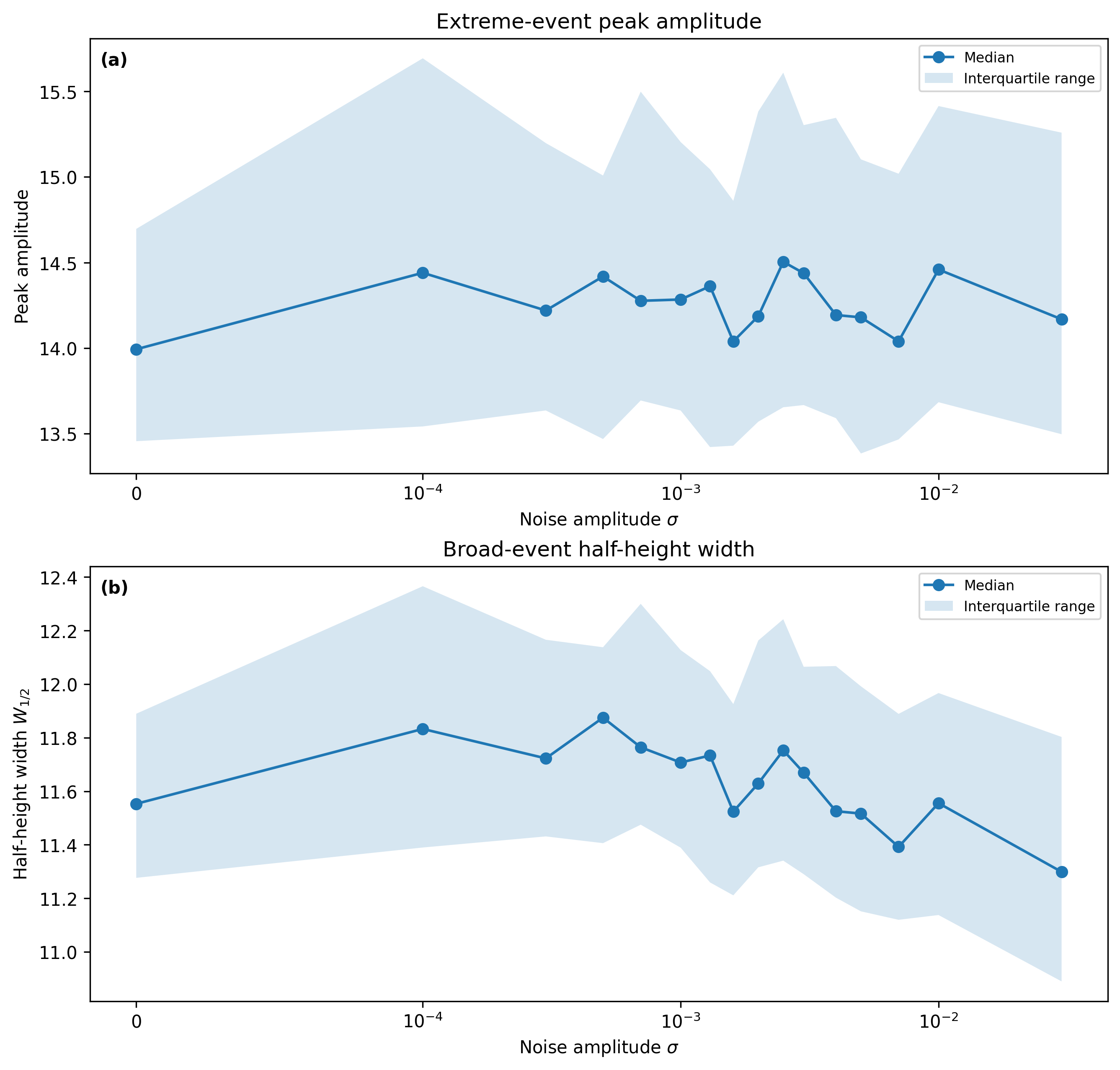}
\caption{Reference dense-scan broad-event morphology. (a) Median EE peak amplitude and interquartile range. (b) Median half-height width $W_{1/2}$ and interquartile range. The morphology varies much less than the occurrence statistics.}
\label{fig:morphology}
\end{figure}

The reference dense scan shows a pronounced change in occurrence statistics (Fig.~\ref{fig:occurrence}). The EE count rises from $45$ in the clean trajectory to $101$ at $\sigma=3\times10^{-2}$, while the median inter-event interval shortens from $501.7$ to $305.4$. In contrast, median amplitude and median $W_{1/2}$ remain within comparatively narrow ranges over the $16$ noise levels (Fig.~\ref{fig:morphology}). This motivates treating noise primarily as a perturbation of EE occurrence rate and timing statistics rather than as a simple amplitude inflation.

\FloatBarrier
\subsection{The recent 15-time-unit history contains the useful clean precursor information}
\label{sec:historyselection}

\begin{figure}[!htbp]
\centering
\includegraphics[width=0.93\linewidth]{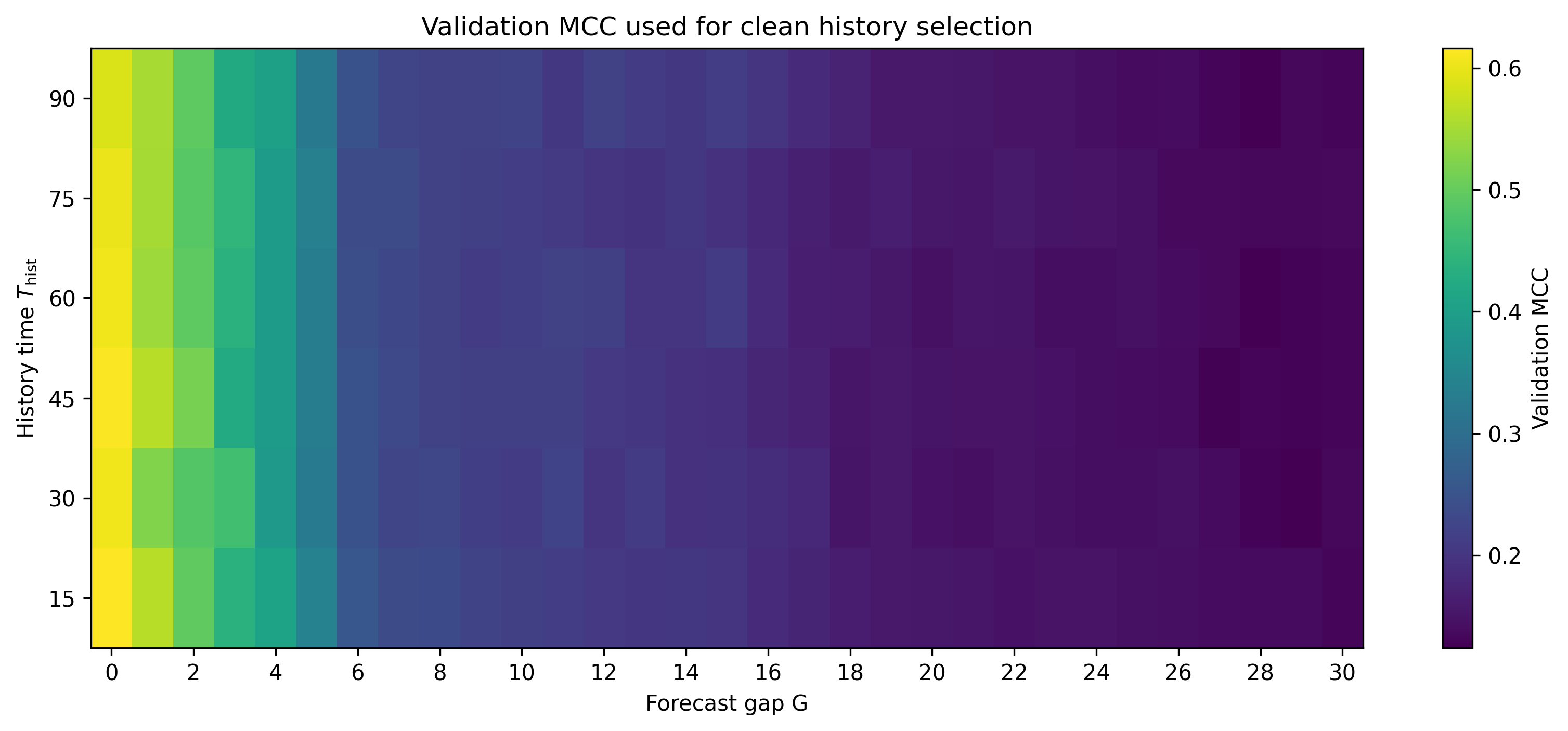}
\caption{Validation MCC used for clean history selection. Extending the history from $15$ to $90$ time units produces no systematic improvement, whereas forecast skill decreases strongly with increasing $G$.}
\label{fig:historyheatmap}
\end{figure}

Figure~\ref{fig:historyheatmap} summarizes validation MCC over all candidate history lengths and forecast gaps. Averaged over $G=0$--$15$, the validation scores are $0.3163$, $0.3096$, $0.3115$, $0.3108$, $0.3088$, and $0.3107$ for $T_{\rm hist}=15$, $30$, $45$, $60$, $75$, and $90$, respectively. The strongest dependence is therefore on forecast gap rather than on older context. The shortest candidate, $T_{\rm hist}=15$, is selected and frozen for the noise experiments.

\subsection{The clean broad-event predictability horizon is short}
\label{sec:cleanhorizon}

With $T_{\rm hist}=15$ fixed from validation, the clean test MCC is $0.641$ at $G=0$, $0.579$ at $G=1$, $0.556$ at $G=2$, $0.483$ at $G=3$, $0.420$ at $G=4$, and $0.369$ at $G=5$. The last gaps satisfying MCC thresholds of $0.5$, $0.4$, $0.3$, and $0.2$ are $G=2$, $4$, $5$, and $6$, respectively.

\begin{figure}[!htbp]
\centering
\includegraphics[width=0.91\linewidth]{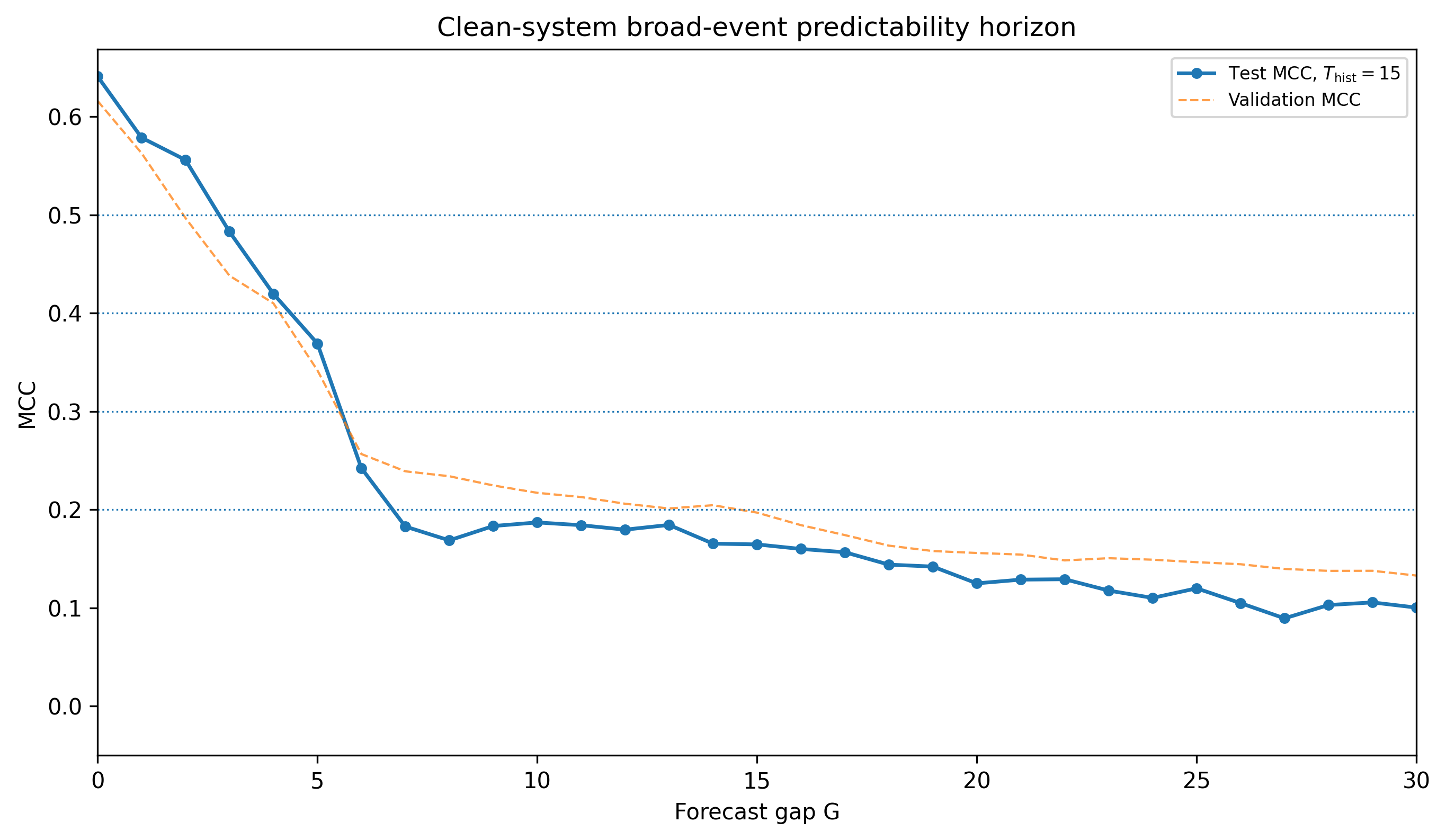}
\caption{Clean-system broad-event predictability horizon for $T_{\rm hist}=15$ and $W=15$. Prediction skill decreases sharply over the first six time units and then forms only a weak residual tail.}
\label{fig:cleanhorizon}
\end{figure}

Beyond the near-event range, MCC remains positive but weak. Specifically, $\mcc(10)=0.187$ and $\mcc(15)=0.165$; at longer gaps, $\mcc(20)=0.125$ and $\mcc(30)=0.100$. The mean test MCC is $0.508$ over $G=0$--$5$, $0.365$ over $G=0$--$10$, and $0.306$ over $G=0$--$15$. The revised broad-event target therefore produces one short high-skill regime followed by a weak tail.

\FloatBarrier
\subsection{Paired ensemble reveals noise-induced predictability redistribution}
\label{sec:ensembleresults}

The paired ten-realization ensemble reveals the defining signature of noise-induced predictability redistribution (NIPR): the response to dynamical noise is non-monotonic in $\sigma$ and nonuniform across forecast gap $G$, rather than a common shift of MCC at every lead time. In the clean paired ensemble,
\begin{equation}
\langle S_{\rm near}\rangle=0.456
\qquad (95\%~{\rm CI}:~0.421\text{--}0.491).
\end{equation}
Very weak noise produces a small decrease, $\langle S_{\rm near}\rangle=0.435$ at $\sigma=5\times10^{-4}$. For intermediate noise, however, the score rises to $0.524$--$0.546$ over $\sigma=0.002$--$0.01$ and reaches its largest ensemble mean at
\begin{equation}
\sigma=7\times10^{-3},\qquad
\langle S_{\rm near}\rangle=0.546.
\end{equation}
The paired difference from clean is $0.0895$ with 95\% CI $0.0494$--$0.1295$, corresponding to $d_z=1.60$; 9 of 10 replicates improve and the Holm-adjusted Wilcoxon value is $p=0.0234$. At $\sigma=10^{-2}$, the mean score is $0.532$, the paired difference is $0.0759$ (95\% CI $0.0437$--$0.1080$), all 10 replicates improve, and the Holm-adjusted $p=0.0137$. Thus the most robust statistical component of NIPR is a reproducible short-horizon gain rather than a universal improvement at all lead times.

\begin{figure}[!htbp]
\centering
\includegraphics[width=0.86\linewidth]{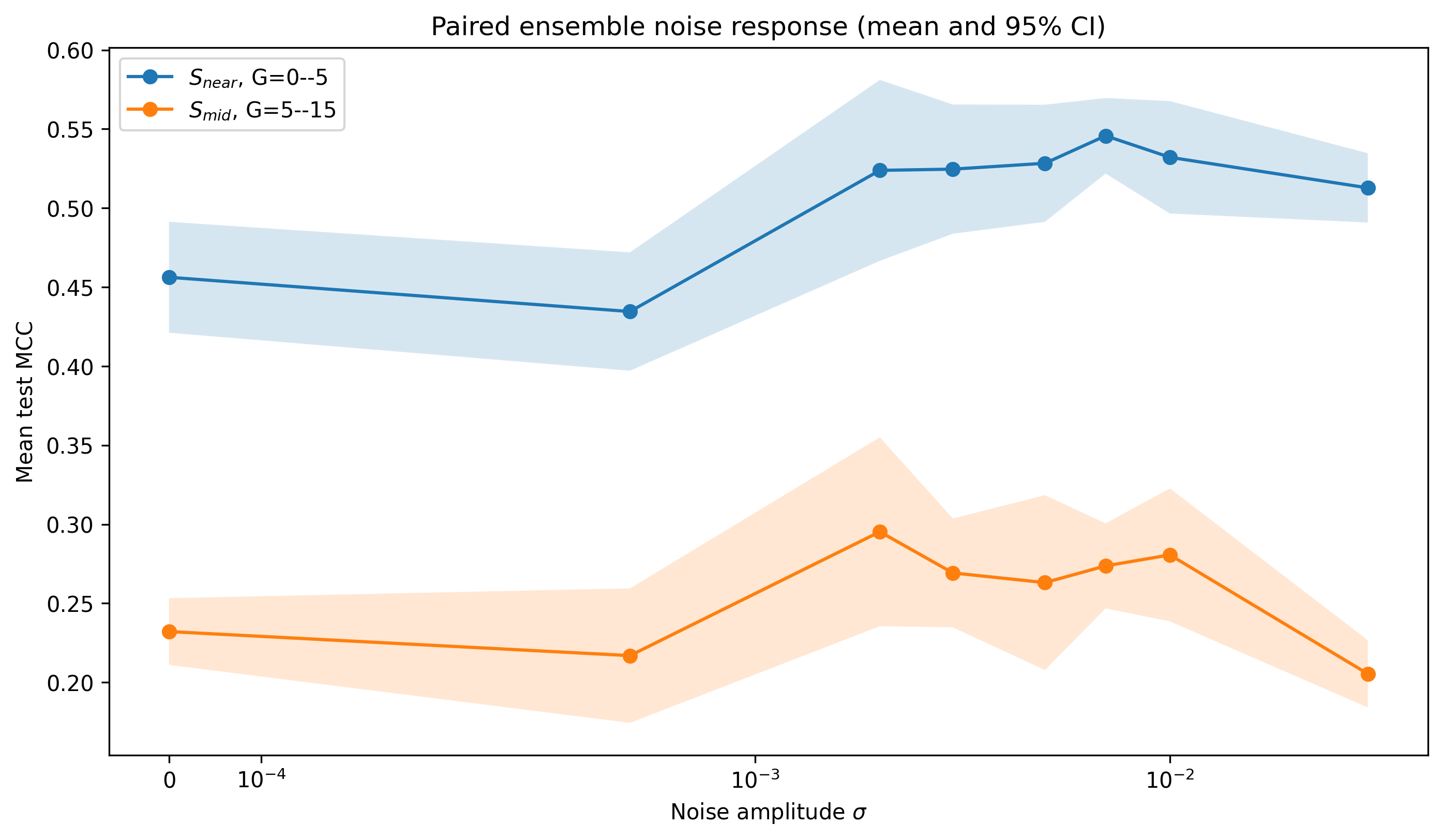}\\[-2pt]
\includegraphics[width=0.86\linewidth]{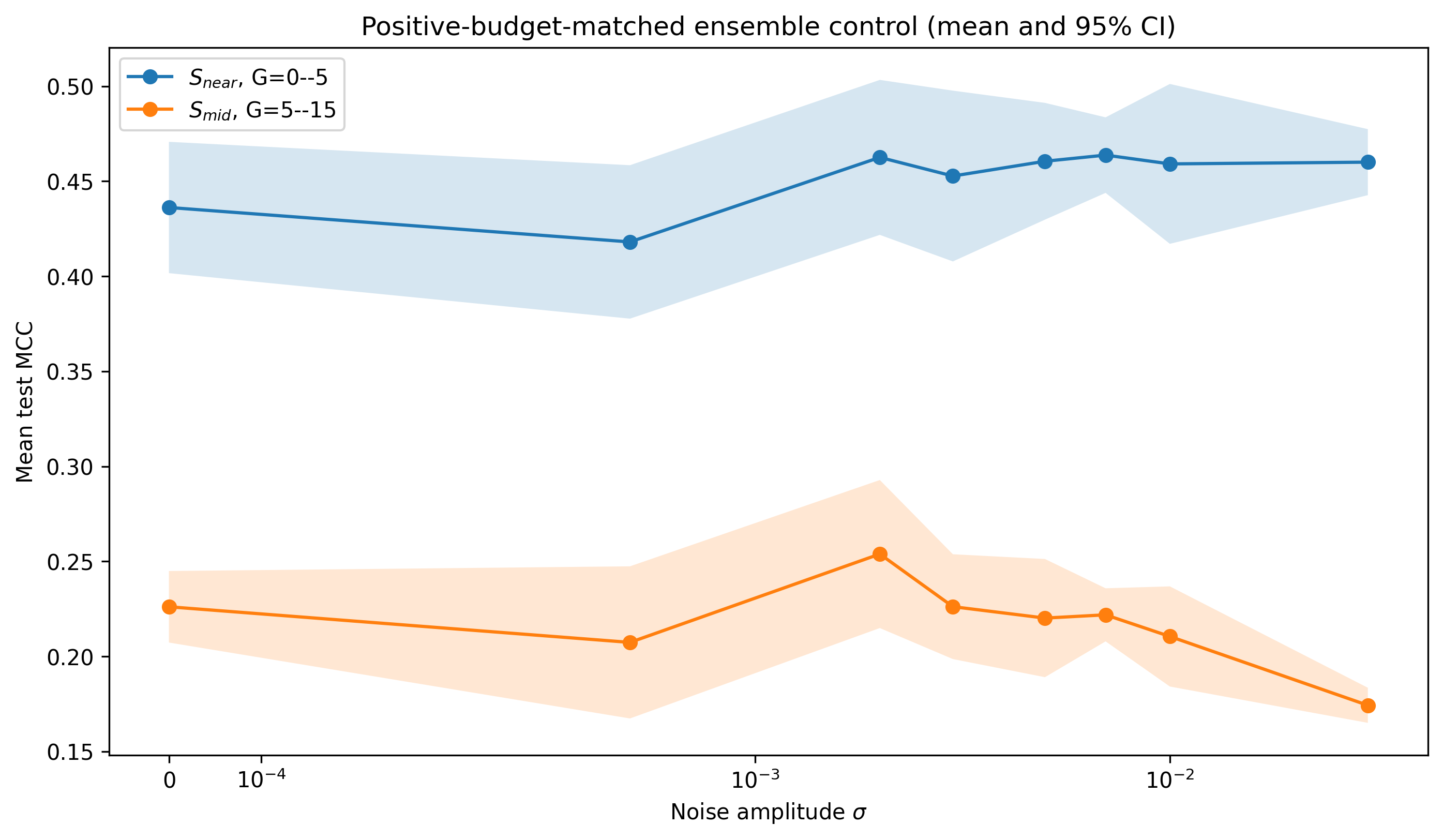}
\caption{Paired ensemble summary of horizon-dependent prediction skill. Top: natural training data. Bottom: positive-training-count-matched control. Curves show ensemble mean $S_{\rm near}$ and $S_{\rm mid}$; shading denotes 95\% confidence intervals over ten paired realizations. Intermediate noise reproducibly improves the natural short-horizon score, whereas matching the positive training count substantially attenuates the gain.}
\label{fig:ensemblescores}
\end{figure}

Figure~\ref{fig:ensemblescores} contrasts the natural-data result with the positive-budget-matched control. The primary response exhibits a broad intermediate-noise enhancement rather than a sharp isolated resonance. The matched analysis substantially compresses the vertical separation between clean and noisy curves.

The intermediate-horizon score is more variable. The clean value is $\langle S_{\rm mid}\rangle=0.232$, while the largest mean, $0.295$, occurs at $\sigma=0.002$. Several intermediate-noise levels have positive paired mean differences and confidence intervals excluding zero, but none remains significant after Holm correction across the seven nonzero noise levels. Consequently, the data support a tendency toward intermediate-horizon improvement but do not establish a unique statistically confirmed optimum for $S_{\rm mid}$.

\begin{figure}[!htbp]
\centering
\includegraphics[width=0.93\linewidth]{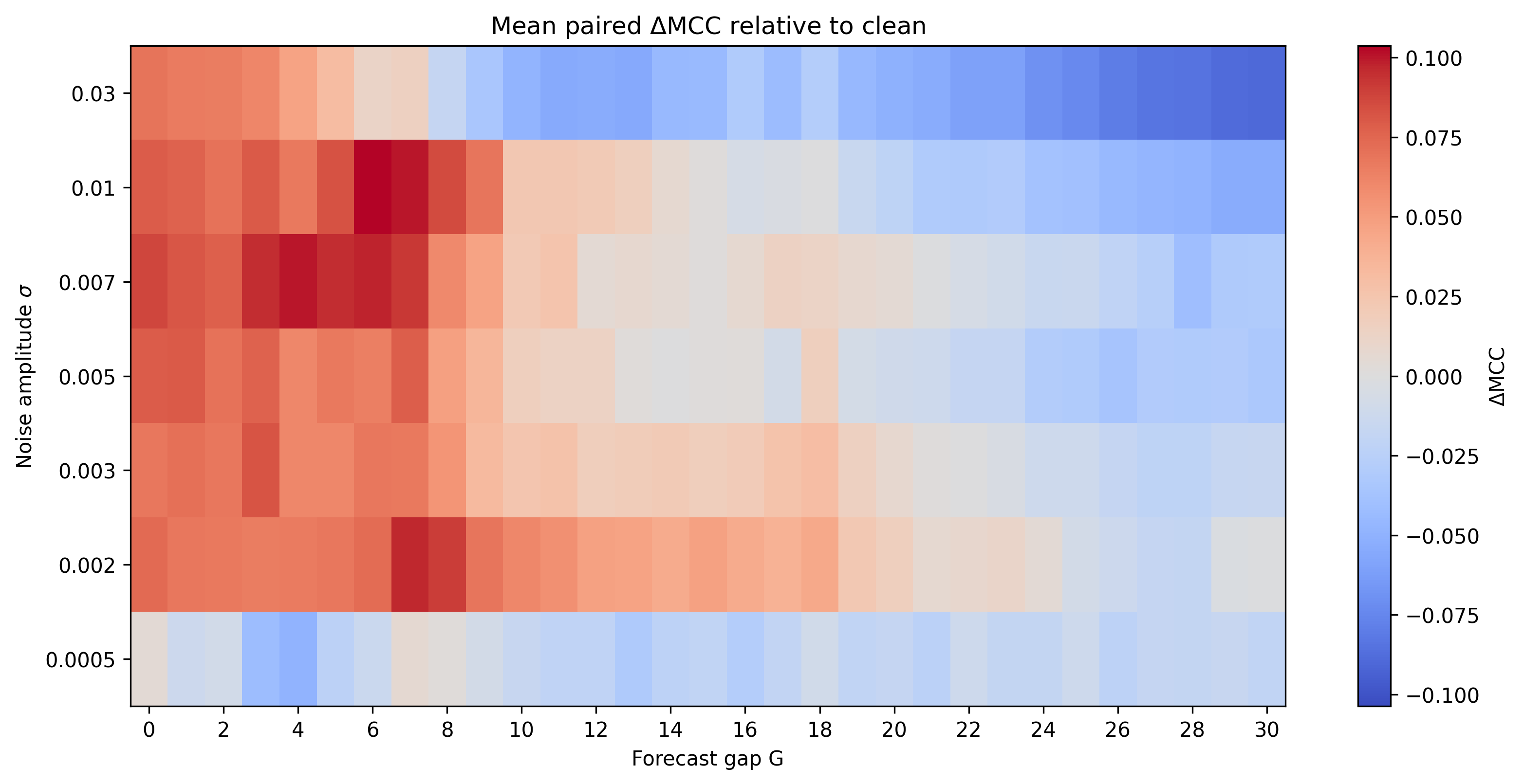}
\caption{Noise-induced predictability redistribution (NIPR) map: mean paired change in test MCC relative to the corresponding clean replicate. Positive values indicate noise-assisted prediction and negative values indicate degradation. The positive region is concentrated at shorter forecast gaps, whereas strong noise increasingly suppresses longer-horizon skill.}
\label{fig:deltaheatmap}
\end{figure}

The redistribution is visible directly in the paired $\Delta\mcc(G,\sigma)$ map (Fig.~\ref{fig:deltaheatmap}), which we therefore refer to as the NIPR map. Intermediate noise produces a positive band concentrated at short and moderate gaps, whereas strong noise progressively removes the advantage at larger $G$. At $\sigma=3\times10^{-2}$, the primary $S_{\rm near}$ remains above the paired clean mean ($0.513$ versus $0.456$), but $S_{\rm mid}$ falls to $0.205$, below the clean value $0.232$. The change in both magnitude and sign of the noise response across $G$ is the operational NIPR signature defined in Eq.~\eqref{eq:niprmap}.

\subsection{Count matching decomposes the NIPR response and exposes strong-noise degradation}

The count-matched control clarifies the origin of the primary enhancement. After forcing every $\sigma$ within a replicate to use the same number of positive training anchors at each $G$, the clean short-horizon score is $0.436$. The largest matched values are $0.464$ at $\sigma=0.007$ and $0.460$ at $\sigma=0.03$, corresponding to paired increases of approximately $0.028$ and $0.024$. None of the positive matched $S_{\rm near}$ differences is significant after Holm correction. The intermediate-horizon enhancement is attenuated even more strongly: matched $\langle S_{\rm mid}\rangle$ is $0.226$ in clean data and $0.254$ at $\sigma=0.002$, but the paired difference is not significant after correction.

At strong noise the matched result reverses sign. For $\sigma=3\times10^{-2}$,
\begin{equation}
\langle S_{\rm mid}^{\rm matched}\rangle=0.174
\end{equation}
compared with $0.226$ in the paired clean ensemble. The paired difference is $-0.0518$ (95\% CI $-0.0715$ to $-0.0322$), all ten replicates decrease, and the Holm-adjusted $p=0.0137$. This provides direct evidence that strong noise destroys information relevant to intermediate-horizon forecasting even though immediate pre-event prediction can remain comparatively strong.

\begin{figure}[!htbp]
\centering
\includegraphics[width=0.86\linewidth]{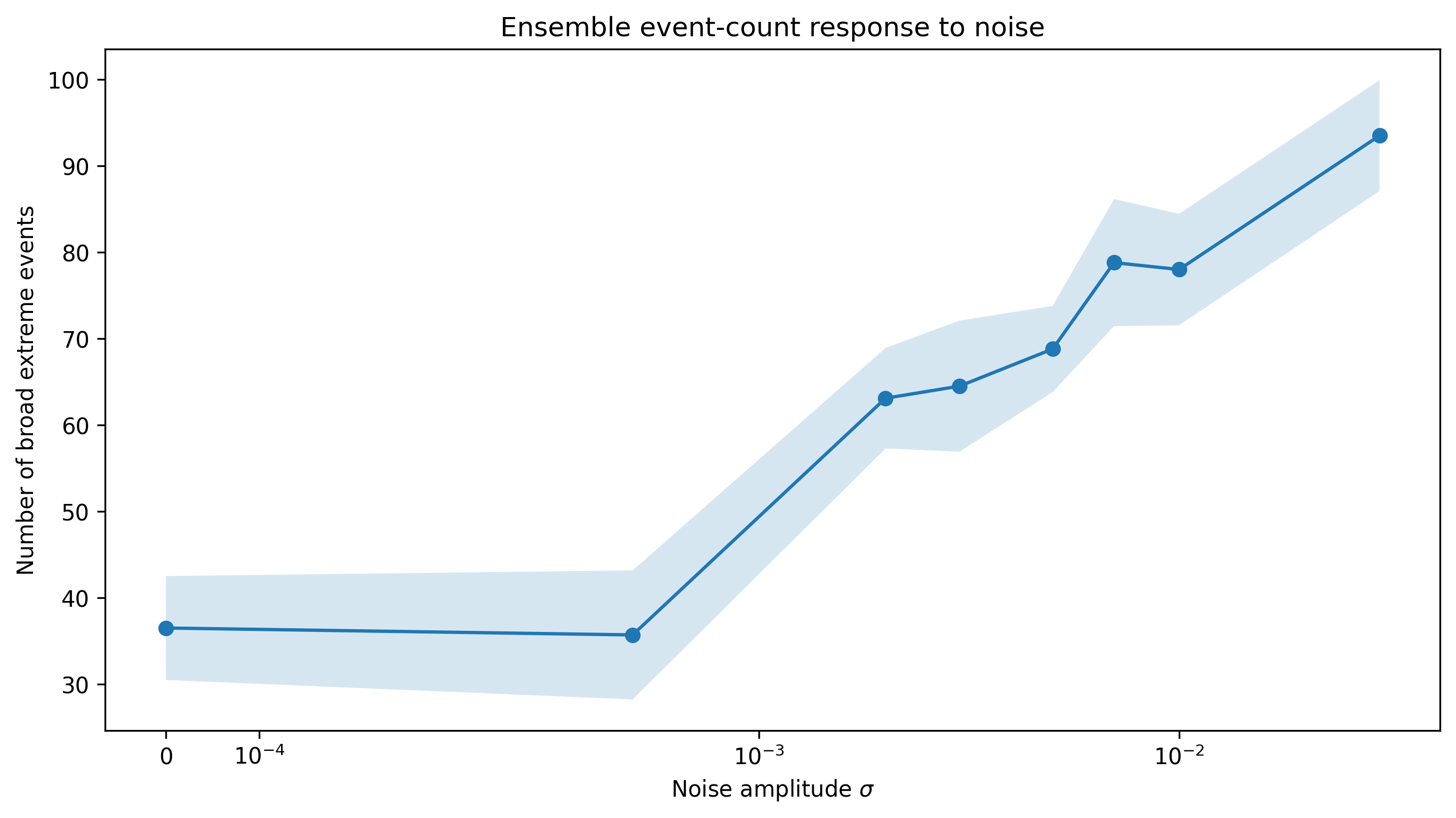}
\caption{Ensemble number of selected broad EEs as a function of dynamical-noise amplitude. Points are means over ten paired realizations and shading denotes 95\% confidence intervals. Noise strongly increases event occurrence while broad-event amplitude and width remain comparatively stable.}
\label{fig:ensembleevents}
\end{figure}

The same ensemble confirms the occurrence/morphology separation suggested by the dense reference scan. The mean number of selected EEs rises from $36.5$ in the perturbed clean ensemble to $63.1$ at $\sigma=0.002$, $78.8$ at $\sigma=0.007$, and $93.5$ at $\sigma=0.03$ (Fig.~\ref{fig:ensembleevents}). Over the same range, the ensemble mean of the realization-level median peak amplitude stays near $14.2$--$14.35$, while the mean median width decreases only modestly from $11.78$ to $11.35$. The mean median inter-event interval shortens from $847.4$ in the paired clean ensemble to $424.6$ at $\sigma=0.007$ and $364.2$ at $\sigma=0.03$.

The count-matched analysis clarifies why prediction improves at intermediate noise levels. Because noise increases the number of EEs, the original noisy training sets contain more positive examples than the clean set. After equalizing the number of positive training examples across noise levels, the qualitative dependence on noise remains similar, but the magnitude of the short-horizon improvement is strongly reduced. At $\sigma=0.007$, for example, the increase in $S_{\rm near}$ relative to clean decreases from about $0.090$ in the primary analysis to about $0.028$ after count matching. Thus the larger number of positive training examples explains a substantial part of the observed short-horizon gain, although a smaller residual noise-dependent increase remains. With ten realizations this residual increase is not statistically significant. In contrast, the deterioration of intermediate-horizon prediction at the strongest noise level remains after count matching, indicating a genuine loss of predictive information at longer lead times.

\section{Discussion: noise-induced predictability redistribution}
\label{sec:discussion}

The revised workflow separates event definition, clean predictability, and noise validation. This distinction is essential for the present system because a broad EE contains internal oscillatory micropeaks. The single frozen threshold $\hee$ supplies one amplitude criterion across all realizations, while Gaussian broad-peak extraction prevents microstructure from being interpreted as multiple physical events. Defining $W_{1/2}$ relative to the realization mean rather than to the threshold further separates event morphology from the event-selection boundary. This event-centered construction is consistent with the broader shift in EE forecasting from generic trajectory extrapolation toward targets tied to specific transient structures and event mechanisms \cite{Farazmand2019,Guth2019,Blonigan2018,Doan2021}.

The clean prediction analysis shows that useful precursor information is concentrated close to the event. Histories longer than $15$ time units do not improve validation MCC, and test skill decreases rapidly with forecast gap: MCC remains above $0.5$ only through $G=2$ and above $0.3$ through $G=5$. The resulting picture is therefore simple: a short high-skill regime is followed by a weak residual tail at larger gaps. Such horizon dependence is consistent with recent work showing both irreducible limits to EE forecasting and strong event-to-event variations in predictability \cite{Yuan2024,VelaMartin2024JFM,VelaMartin2024PRF}. Very recent studies further emphasize event-specific predictability hierarchies and dynamics-informed precursors \cite{Yang2026,Katsidoniotaki2026,Consonni2026}.

The paired ensemble adds the central result of this study: noise does not shift prediction skill uniformly, but redistributes it across forecast horizons. Very weak perturbations slightly reduce $S_{\rm near}$, intermediate noise increases it, and stronger noise confines any advantage increasingly to the immediate pre-event interval while degrading more distant skill. We identify this pattern as the \emph{noise-induced predictability redistribution} (NIPR) effect. The strongest reproducible positive component occurs at short horizons for $\sigma=7\times10^{-3}$ and $10^{-2}$, where paired improvements remain significant after Holm correction; the negative component becomes clear at strong noise in the matched intermediate-horizon score.

The event statistics and the count-matched control make the physical interpretation more transparent. Noise substantially increases the number of broad EEs and shortens their typical temporal separation, while peak amplitude and width change much less. Noise-induced transition theory provides a natural context for such behavior: stochastic forcing can modify escape rates, first-passage times, attractor hopping, recurrence, and extreme-event clustering \cite{Forgoston2017,Faranda2012,Huang2022,Zhao2023}. The direction is not universal, because noise can also suppress extreme events in other nonlinear systems \cite{Sen2021}. Thus the increase in EE density observed here is a property of this dynamical regime rather than a general consequence of adding noise.

This increase in event number also creates a straightforward machine-learning advantage: the natural noisy training sets contain more positive precursor examples. The count-matched control removes this advantage by forcing every noise level to use the same number of positive training examples, while leaving validation and test data unchanged. After this correction, the shape of the noise dependence remains qualitatively similar, but most of the short-horizon gain is reduced and the remaining positive differences are not statistically significant with ten realizations. We therefore do not claim that intermediate noise has been proven to intrinsically sharpen deterministic precursors. What is supported is that dynamical noise reorganizes the EE process, and this reorganization changes forecastability; part of the observed short-horizon improvement is specifically explained by the increased availability of positive training examples. At strong noise, the matched intermediate-horizon degradation remains significant, showing that the noise can also remove predictive information at longer lead times.

The non-monotonic dependence of forecast skill on noise amplitude is qualitatively reminiscent of resonance-like phenomena: intermediate noise improves short-horizon prediction, whereas stronger noise reduces skill at longer horizons. However, the present results do not establish a stochastic- or coherence-resonance mechanism. Coherence resonance concerns noise-induced temporal organization \cite{Pikovsky1997}, and Revelli \emph{et al.} reported stochastic-resonance-like forecasting behavior under external noise \cite{Revelli2010}; training-noise effects can also improve machine-learning forecasts through regularization \cite{Zhai2023,Wikner2023,Vlachas2018}. In our data, shorter inter-event intervals show that EEs occur more frequently, but frequency alone does not demonstrate that event timing becomes more regular or that a particular spectral component is amplified. Establishing such a mechanism would require additional evidence, for example a noise-dependent reduction in the coefficient of variation of inter-event intervals, enhanced autocorrelation, or the emergence of a distinct spectral peak. We therefore use NIPR as the more conservative description: the noise acts directly on the physical evolution, reorganizes the EE process, and changes forecast skill differently across lead times. To the best of our knowledge, this horizon-resolved redistribution of EE predictability under dynamical noise has not previously been formulated explicitly as NIPR.

Strong noise exposes the second side of NIPR. At $\sigma=3\times10^{-2}$, matched $S_{\rm mid}$ decreases in all ten replicates and remains significant after Holm correction. Thus the effect is not merely a positive short-horizon enhancement: it includes a noise-induced loss of farther-ahead information, so the predictive response changes character across lead time. The NIPR map captures this directly without forcing the response into a single scalar ``optimal noise'' value. This target- and horizon-dependent interpretation echoes earlier findings that stochastic forcing can influence trajectory, regime, and climate predictability in different ways \cite{Kwasniok2014,Sardeshmukh2023,Sangiorgio2021Forecasting}, while extending the analysis to the explicit future-window forecasting of broad EEs.

The paired design improves efficiency by comparing the same tiny initial-condition perturbation and the same underlying Gaussian-noise path across $\sigma$ within a replicate. Nevertheless, several limitations should be noted. The ensemble contains ten realizations, the representative $\sigma$ subset was chosen after an exploratory dense scan rather than preregistered independently, and the analysis is restricted to one chaotic system and one observable. The probability operating threshold is selected separately on validation data for each realization, $\sigma$, and $G$, which is appropriate for operational classification but means the reported MCC describes optimized alarm operation rather than calibration transfer across noise regimes. The literature also shows that observation noise, training noise, and dynamical forcing can have qualitatively different forecasting effects \cite{Sangiorgio2021Sensitivity,Vlachas2018,Zhai2023}; conclusions from one setting should therefore not be transferred automatically to another. Future work should test whether the same horizon-dependent pattern persists in other EE-generating systems, with alternative classifiers, and under experimentally measured rather than synthetic noise. An important next step is to resolve NIPR in phase space and determine which precursory regions gain or lose predictive information as the noise amplitude changes.

Overall, the simplest interpretation is that dynamical noise changes the EE-generating process itself and, as a consequence, changes how predictable the events are at different lead times. Broad-event geometry remains comparatively stable, but event occurrence is strongly reorganized and forecast skill responds nonuniformly across $G$. The count-matched experiment shows that the increased number of positive training examples explains a substantial part of the natural short-horizon improvement, while the surviving strong-noise degradation demonstrates that the response is not only a training-sample-count effect. NIPR therefore describes this combined dynamical and forecasting change without requiring a claim of classical stochastic resonance.

\section{Conclusions}
\label{sec:conclusions}

We developed a single-threshold broad-event framework for forecasting extreme excursions in a chaotic flow and used it to distinguish event geometry, clean predictability horizon, and noise-dependent forecast skill. The fixed threshold $\hee=13.052073$ is computed once from the reference clean peak distribution and retained for every subsequent realization. One coordinate is assigned to each broad excursion, exact peak amplitude is recovered from the raw signal, and width is measured at half height relative to the realization mean. The reference clean median $W_{1/2}=11.553$ motivates the fixed future window $W=15$.

Chronological validation selects $T_{\rm hist}=15$; older context up to $90$ time units provides no systematic benefit. The clean test MCC decreases from $0.641$ at $G=0$ to $0.369$ at $G=5$ and $0.165$ at $G=15$, establishing a short useful predictability horizon followed by a weak residual tail.

The final paired ten-realization ensemble shows that dynamical noise produces a non-monotonic and horizon-dependent response. The natural-data short-horizon score increases from $\langle S_{\rm near}\rangle=0.456$ in the paired clean ensemble to $0.546$ at $\sigma=7\times10^{-3}$, a paired increase of $0.0895$ with 95\% CI $0.0494$--$0.1295$ and Holm-adjusted $p=0.0234$. At $\sigma=10^{-2}$ all ten paired realizations improve and the adjusted $p=0.0137$. Intermediate-horizon enhancement is more variable and is not significant after multiplicity correction.

Noise also strongly increases EE occurrence while leaving broad-event amplitude and width comparatively stable. When the number of positive training anchors is equalized across noise levels, most of the short-horizon gain is attenuated and the remaining positive differences are not statistically significant. Conversely, at $\sigma=3\times10^{-2}$ the matched intermediate-horizon score decreases by $0.0518$ relative to paired clean data, with all ten replicates decreasing and Holm-adjusted $p=0.0137$.

These results identify and characterize a \emph{noise-induced predictability redistribution} (NIPR) effect, a term introduced here for the nonuniform change of EE forecast skill across lead times under dynamical noise, rather than a universal increase or decrease in predictability. Intermediate noise produces the reproducible positive NIPR component near the event, partly through increased event density and positive-example availability, whereas strong noise produces a negative component at longer horizons. The count-matched control shows that most of the natural short-horizon gain is coupled to noise-induced reorganization of the EE process; any residual intrinsic enhancement is smaller and not statistically resolved here. NIPR therefore provides a compact description of the joint dependence of extreme-event forecastability on noise intensity and forecast horizon without requiring a classical stochastic-resonance interpretation.

\section*{CRediT authorship contribution statement}
\textbf{Andrei Velichko:} Conceptualization, Methodology, Software, Formal analysis, Investigation, Data curation, Visualization, Writing---original draft preparation, Writing---review and editing, Funding acquisition. \textbf{Viet-Thanh Pham:} Conceptualization, Validation, Investigation, Resources, Supervision, Project administration, Writing---review and editing. All authors have read and agreed to the published version of the manuscript.

\section*{Declaration of competing interest}
The authors declare that they have no known competing financial interests or personal relationships that could have appeared to influence the work reported in this paper.

\section*{Data and code availability}
The numerical workflow is organized as reproducible Supplementary Programs 01--06. Programs 01--03 generate the reference clean/noisy trajectories, construct the single-threshold broad-event catalogs, and produce event-statistics figures. Program 04 performs clean history selection and predictability-horizon analysis. Program 05 provides the exploratory dense noise-prediction scan. Program 06 performs the compact paired ten-realization ensemble validation, including the positive-training-count-matched control and the final mean/CI and paired-test tables. The ensemble program regenerates trajectories in memory and stores only compact event, prediction, and statistical outputs.

\section*{Funding}
This research was funded by the Russian Science Foundation, Grant No. 22-11-00055-P.

\bibliographystyle{unsrt}
\bibliography{references}

@article{Durairaj2026,
  author  = {Premraj Durairaj and Thamilmaran Kathamuthu and Zhigang Zheng},
  title   = {Unveiling Extreme Events in a Nonlinear Chaotic System: Numerical and Experimental Realizations},
  journal = {International Journal of Bifurcation and Chaos},
  volume  = {36},
  number  = {4},
  pages   = {2630008},
  year    = {2026},
  doi     = {10.1142/S0218127426300089}
}

@article{LiSprott2016,
  author  = {Chunbiao Li and Julien Clinton Sprott},
  title   = {Variable-boostable chaotic flows},
  journal = {Optik},
  volume  = {127},
  pages   = {10389--10398},
  year    = {2016}
}

@article{Chowdhury2022,
  author  = {Subhadeep Nag Chowdhury and Avijit Ray and Sudeshna K. Dana and Dibakar Ghosh},
  title   = {Extreme events in dynamical systems and random walkers: A review},
  journal = {Physics Reports},
  volume  = {966},
  pages   = {1--52},
  year    = {2022}
}

@article{Mishra2020,
  author  = {Awadhesh Mishra and S. Leo Kingston and C. Hens and Tomasz Kapitaniak and Ulrike Feudel and Sudeshna K. Dana},
  title   = {Routes to extreme events in dynamical systems: Dynamical and statistical characteristics},
  journal = {Chaos},
  volume  = {30},
  pages   = {063114},
  year    = {2020}
}

@article{Cavalcante2013,
  author  = {Hugo L. D. de S. Cavalcante and M. Oria and Didier Sornette and Edward Ott and Daniel J. Gauthier},
  title   = {Predictability and control of extreme events in complex systems},
  journal = {Physical Review Letters},
  volume  = {111},
  pages   = {198701},
  year    = {2013}
}

@article{Bialonski2015,
  author  = {Stephan Bialonski and Gerrit Ansmann and Holger Kantz},
  title   = {Data-driven prediction and prevention of extreme events in a spatially extended excitable system},
  journal = {Physical Review E},
  volume  = {92},
  pages   = {042910},
  year    = {2015}
}

@article{ChenMajda2020,
  author  = {Nan Chen and Andrew J. Majda},
  title   = {Predicting observed and hidden extreme events in complex nonlinear dynamical systems with partial observations and short training time series},
  journal = {Chaos},
  volume  = {30},
  pages   = {033101},
  year    = {2020}
}

@article{Farazmand2019,
  author  = {Farazmand, M. and Sapsis, T.},
  title   = {Extreme Events: Mechanisms and Prediction},
  journal = {Applied Mechanics Reviews},
  volume  = {71},
  pages   = {050801 },
  year    = {2019},
  doi     = {10.1115/1.4042065}
}

@article{Guth2019,
  author  = {Guth, S. and Sapsis, T.},
  title   = {Machine Learning Predictors of Extreme Events Occurring in Complex Dynamical Systems},
  journal = {Entropy},
  volume  = {21},
  year    = {2019},
  doi     = {10.3390/e21100925}
}

@article{Blonigan2018,
  author  = {Blonigan, Patrick J. and Farazmand, M. and Sapsis, T.},
  title   = {Are extreme dissipation events predictable in turbulent fluid flows?},
  journal = {Physical Review Fluids},
  year    = {2018},
  doi     = {10.1103/physrevfluids.4.044606}
}

@article{Doan2021,
  author  = {Doan, N. and Polifke, W. and Magri, L.},
  title   = {Short- and long-term predictions of chaotic flows and extreme events: a physics-constrained reservoir computing approach},
  journal = {Proceedings of the Royal Society A},
  volume  = {477},
  year    = {2021},
  doi     = {10.1098/rspa.2021.0135}
}

@article{Pammi2023,
  author  = {Pammi, V. and Clerc, M. and Coulibaly, S. and Barbay, S.},
  title   = {Extreme Events Prediction from Nonlocal Partial Information in a Spatiotemporally Chaotic Microcavity Laser},
  journal = {Physical Review Letters},
  volume  = {130},
  pages   = {223801},
  year    = {2023},
  doi     = {10.1103/physrevlett.130.223801}
}

@article{Ahmed2024,
  author  = {Ahmed, Osama and Tennie, Felix and Magri, Luca},
  title   = {Prediction of chaotic dynamics and extreme events: A recurrence-free quantum reservoir computing approach},
  journal = {Physical Review Research},
  year    = {2024},
  doi     = {10.1103/physrevresearch.6.043082}
}

@article{Yuan2024,
  author  = {Yuan, Yuan and Lozano-Duran, A.},
  title   = {Limits to extreme event forecasting in chaotic systems},
  journal = {Physica D: Nonlinear Phenomena},
  year    = {2024},
  doi     = {10.1016/j.physd.2024.134246}
}

@article{VelaMartin2024JFM,
  author  = {Vela-Martin, A. and Avila, Marc},
  title   = {Large-scale patterns set the predictability limit of extreme events in Kolmogorov flow},
  journal = {Journal of Fluid Mechanics},
  volume  = {986},
  year    = {2024},
  doi     = {10.1017/jfm.2024.263}
}

@article{VelaMartin2024PRF,
  author  = {Vela-Martin, A.},
  title   = {Complexity of extreme-event prediction in turbulent flows},
  journal = {Physical Review Fluids},
  year    = {2024},
  doi     = {10.1103/physrevfluids.9.104603}
}

@article{Yang2026,
  author  = {Yang, Yuxuan and Dong, Chenyu and Mengaldo, Gianmarco},
  title   = {Hierarchy of extreme-event predictability in turbulence revealed by machine learning},
  journal = {arXiv},
  year    = {2026},
  doi     = {10.48550/arxiv.2603.13789}
}

@article{Katsidoniotaki2026,
  author  = {Katsidoniotaki, Eirini and Sapsis, Themistoklis P.},
  title   = {Dynamics-Informed Deep Learning for Predicting Extreme Events},
  journal = {arXiv},
  year    = {2026},
  doi     = {10.48550/arxiv.2603.10777}
}

@article{Consonni2026,
  author  = {Consonni, Riccardo and Magri, Luca},
  title   = {Precursors of extreme events and critical transitions},
  journal = {arXiv},
  year    = {2026},
  doi     = {10.48550/arxiv.2604.12869}
}

@article{Martin2025,
  author  = {Martin, David and Grau, Joan and Jofre, L.},
  title   = {Conditional POD for predicting extreme events in turbulent flow time signals},
  journal = {Scientific Reports},
  volume  = {15},
  year    = {2025},
  doi     = {10.1038/s41598-025-14804-4}
}

@article{Wang2024,
  author  = {Wang, Tao and Zhou, Hanxu and Fang, Qing and Han, Yanan and Guo, Xingxing and Zhang, Yahui and Qian, Chao and Chen, Hongsheng and Barland, Stephane and Xiang, S. and Lippi, G. L.},
  title   = {Reservoir computing-based advance warning of extreme events},
  journal = {Chaos, Solitons \& Fractals},
  year    = {2024},
  doi     = {10.1016/j.chaos.2024.114673}
}

@article{Pickering2022,
  author  = {Pickering, Ethan and Guth, Stephen and Karniadakis, George Em and Sapsis, Themistoklis P.},
  title   = {Discovering and forecasting extreme events via active learning in neural operators},
  journal = {Nature Computational Science},
  volume  = {2},
  pages   = {823--833},
  year    = {2022},
  doi     = {10.1038/s43588-022-00376-0}
}

@article{Mishra2022,
  author  = {Mishra, Arindam and Dana, S. K. and Hens, C. and Kapitaniak, T. and Kurths, Jurgen and Marwan, N.},
  title   = {Predicting the data structure prior to extreme events from passive observables using echo state network},
  journal = {Frontiers in Applied Mathematics and Statistics},
  volume  = {8},
  year    = {2022},
  doi     = {10.3389/fams.2022.955044}
}

@article{Meiyazhagan2021,
  author  = {Meiyazhagan, J. and Sudharsan, S. and Venkatesan, A. and Senthilvelan, M.},
  title   = {Prediction of occurrence of extreme events using machine learning},
  journal = {European Physical Journal Plus},
  volume  = {137},
  year    = {2021},
  doi     = {10.1140/epjp/s13360-021-02249-3}
}

@article{Farmer1987,
  author  = {Farmer, J. and Sidorowich, J. J.},
  title   = {Predicting chaotic time series},
  journal = {Physical Review Letters},
  volume  = {59},
  pages   = {845--848},
  year    = {1987},
  doi     = {10.1103/physrevlett.59.845}
}

@article{Platzer2020,
  author  = {Platzer, P. and Yiou, P. and Naveau, P. and Tandeo, P. and Zhen, Y. and Ailliot, P. and Filipot, J.},
  title   = {Using local dynamics to explain analog forecasting of chaotic systems},
  journal = {Journal of the Atmospheric Sciences},
  year    = {2020},
  doi     = {10.1175/jas-d-20-0204.1}
}

@article{Zhai2023,
  author  = {Zhai, Zheng-Meng and Kong, Ling-Wei and Lai, Ying-Cheng},
  title   = {Emergence of a resonance in machine learning},
  journal = {Physical Review Research},
  year    = {2023},
  doi     = {10.1103/physrevresearch.5.033127}
}

@article{Wikner2023,
  author  = {Wikner, Alexander and Harvey, Joseph and Girvan, M. and Hunt, Brian R. and Pomerance, A. and Antonsen, Thomas and Ott, Edward},
  title   = {Stabilizing machine learning prediction of dynamics: Novel noise-inspired regularization tested with reservoir computing},
  journal = {Neural Networks},
  volume  = {170},
  pages   = {94--110},
  year    = {2023},
  doi     = {10.1016/j.neunet.2023.10.054}
}

@article{Kwasniok2014,
  author  = {Kwasniok, F.},
  title   = {Enhanced regime predictability in atmospheric low-order models due to stochastic forcing},
  journal = {Philosophical Transactions of the Royal Society A},
  volume  = {372},
  year    = {2014},
  doi     = {10.1098/rsta.2013.0286}
}

@article{Revelli2010,
  author  = {Revelli, J. and Rodriguez, Miguel A. and Wio, H.},
  title   = {Interplay between Chaos and External Noise in an Extended System: Improved Forecasting due to Intrinsic Stochastic Resonant Phenomena},
  journal = {International Journal of Bifurcation and Chaos},
  volume  = {20},
  pages   = {213--224},
  year    = {2010},
  doi     = {10.1142/s021812741002565x}
}

@article{Vlachas2018,
  author  = {Vlachas, Pantelis R. and Byeon, Wonmin and Wan, Z. Y. and Sapsis, T. and Koumoutsakos, P.},
  title   = {Data-driven forecasting of high-dimensional chaotic systems with long short-term memory networks},
  journal = {Proceedings of the Royal Society A},
  volume  = {474},
  year    = {2018},
  doi     = {10.1098/rspa.2017.0844}
}

@article{Sardeshmukh2023,
  author  = {Sardeshmukh, P. and Wang, J. and Compo, G. and Penland, C.},
  title   = {Improving Atmospheric Models by Accounting for Chaotic Physics},
  journal = {Journal of Climate},
  year    = {2023},
  doi     = {10.1175/jcli-d-22-0880.1}
}

@article{Sangiorgio2021Forecasting,
  author  = {Sangiorgio, M. and Dercole, F. and Guariso, G.},
  title   = {Forecasting of noisy chaotic systems with deep neural networks},
  journal = {Chaos, Solitons \& Fractals},
  year    = {2021},
  doi     = {10.1016/j.chaos.2021.111570}
}

@article{Sangiorgio2021Sensitivity,
  author  = {Sangiorgio, M. and Dercole, F. and Guariso, G.},
  title   = {Sensitivity of Chaotic Dynamics Prediction to Observation Noise},
  journal = {IFAC-PapersOnLine},
  year    = {2021},
  doi     = {10.1016/j.ifacol.2021.11.037}
}

@article{Pikovsky1997,
  author  = {Pikovsky, A. and Kurths, J.},
  title   = {Coherence Resonance in a Noise-Driven Excitable System},
  journal = {Physical Review Letters},
  volume  = {78},
  pages   = {775--778},
  year    = {1997},
  doi     = {10.1103/physrevlett.78.775}
}

@article{Forgoston2017,
  author  = {Forgoston, Eric and Moore, Richard O.},
  title   = {A Primer on Noise-Induced Transitions in Applied Dynamical Systems},
  journal = {SIAM Review},
  volume  = {60},
  pages   = {969--1009},
  year    = {2017},
  doi     = {10.1137/17m1142028}
}

@article{Faranda2012,
  author  = {Faranda, Davide and Freitas, J. and Lucarini, V. and Turchetti, G. and Vaienti, S.},
  title   = {Extreme value statistics for dynamical systems with noise},
  journal = {Nonlinearity},
  volume  = {26},
  pages   = {2597--2622},
  year    = {2012},
  doi     = {10.1088/0951-7715/26/9/2597}
}

@article{Huang2022,
  author  = {Huang, Yong},
  title   = {Nonlinear stochastic dynamics research on a Lorenz system with white Gaussian noise based on a quasi-potential approach},
  journal = {International Journal of Mechanical System Dynamics},
  volume  = {3},
  pages   = {85--94},
  year    = {2022},
  doi     = {10.1002/msd2.12062}
}

@article{Zhao2023,
  author  = {Zhao, Dan and Li, Yongge and Liu, Qi and Zhang, Huikang and Xu, Yong},
  title   = {The occurrence mechanisms of extreme events in a class of nonlinear Duffing-type systems under random excitations},
  journal = {Chaos},
  volume  = {33},
  pages   = {083109},
  year    = {2023},
  doi     = {10.1063/5.0156492}
}

@article{Hariharan2025,
  author  = {Hariharan, S. and Suresh, R. and Chandrasekar, V. K.},
  title   = {Noise-induced extreme events in single FitzHugh--Nagumo oscillator},
  journal = {Chaos, Solitons \& Fractals},
  volume  = {192},
  pages   = {116077},
  year    = {2025},
  doi     = {10.1016/j.chaos.2025.116077}
}

@article{Boaretto2025,
  author  = {Boaretto, B. and Macau, E. and Masoller, C.},
  title   = {Noise-induced extreme events in Hodgkin--Huxley neural networks},
  journal = {Chaos, Solitons \& Fractals},
  volume  = {194},
  pages   = {116133},
  year    = {2025},
  doi     = {10.1016/j.chaos.2025.116133}
}

@article{Sen2021,
  author  = {Sen, Deeptajyoti and Sinha, Sudeshna},
  title   = {Enhancement of extreme events through the Allee effect and its mitigation through noise in a three species system},
  journal = {Scientific Reports},
  volume  = {11},
  pages   = {20913},
  year    = {2021},
  doi     = {10.1038/s41598-021-00174-0}
}

\end{document}